\documentclass{article}

\usepackage[preprint]{neurips_2025}

\usepackage[utf8]{inputenc} 
\usepackage[T1]{fontenc}    
\usepackage{hyperref}       
\usepackage{url}            
\usepackage{booktabs}       
\usepackage{amsfonts}       
\usepackage{nicefrac}       
\usepackage{microtype}      
\usepackage{xcolor}         

\usepackage{ragged2e}
\usepackage{booktabs,makecell,multirow,tabularx}
\usepackage{caption}
\usepackage{verbatim}
\usepackage{graphicx} 
\usepackage{float}
\usepackage{subfigure} 
\usepackage{amsmath}

\usepackage{algorithm}
\usepackage{algpseudocode}
\usepackage{setspace}

\title{MotionRAG-Diff: A Retrieval-Augmented Diffusion Framework for Long-Term Music-to-Dance Generation}

\author{
  Mingyang Huang, Peng Zhang, Bang Zhang \\
  Tongyi Lab, Alibaba Group \\
  \texttt{\{hongcan.hmy, futian.zp, zhangbang.zb\}}@alibaba-inc.com
}

\begin{document}

\maketitle

\begin{abstract}
    Generating long-term, coherent, and realistic music-conditioned dance sequences remains a challenging task in human motion synthesis. Existing approaches exhibit critical limitations: motion graph methods rely on fixed template libraries, restricting creative generation; diffusion models, while capable of producing novel motions, often lack temporal coherence and musical alignment. To address these challenges, we propose $\textbf{MotionRAG-Diff}$, a hybrid framework that integrates Retrieval-Augmented Generation (RAG) with diffusion-based refinement to enable high-quality, musically coherent dance generation for arbitrary long-term music inputs. Our method introduces three core innovations: (1) A cross-modal contrastive learning architecture that aligns heterogeneous music and dance representations in a shared latent space, establishing unsupervised semantic correspondence without paired data; (2) An optimized motion graph system for efficient retrieval and seamless concatenation of motion segments, ensuring realism and temporal coherence across long sequences; (3) A multi-condition diffusion model that jointly conditions on raw music signals and contrastive features to enhance motion quality and global synchronization. Extensive experiments demonstrate that MotionRAG-Diff achieves state-of-the-art performance in motion quality, diversity, and music-motion synchronization accuracy. This work establishes a new paradigm for music-driven dance generation by synergizing retrieval-based template fidelity with diffusion-based creative enhancement.
\end{abstract}

\section{Introduction}
\label{section: introduction}
Dance motion generation from music~\cite{ofli2011learn2dance}~\cite{alemi2017groovenet}~\cite{huang2020dance_DanceRevolution}~\cite{li2022danceformer}~\cite{li2021aiFACT_aist++}~\cite{siyao2023bailando++}~\cite{tang2018dance_dance_with_melody}~\cite{tseng2023edge}~\cite{li2024lodge++}~\cite{huang2024beat-it} has emerged as a pivotal research area in human motion synthesis, with significant applications in entertainment, virtual reality, and human-computer interaction. Current approaches predominantly follow two paradigms: motion graph methods~\cite{liu2024tango}~\cite{chen2021choreomaster} that rely on template-based action retrieval and pure generative models like diffusion-based frameworks~\cite{tseng2023edge}~\cite{li2024lodge}~\cite{huang2024beat-it}. However, these approaches exhibit inherent limitations that hinder the creation of high-quality, musically coherent dance sequences. Motion graph methods, while ensuring temporal coherence through pre-defined motion templates, suffer from a fundamental deficiency - their inability to generate novel dance patterns beyond the template library. Conversely, pure diffusion models demonstrate strong generative capabilities but often produce unnatural motion sequences that lag behind the quality of template-based actions. This dichotomy between template fidelity and creative generation remains a critical challenge in the field.

This paper presents a novel hybrid framework that synergistically combines the strengths of motion graphs and diffusion models while addressing their limitations through innovative architectural design. Our key contribution lies in developing a contrastive learning framework that effectively captures the complex correlations between musical features and corresponding dance movements. By integrating this contrastive learning mechanism with an optimized motion graph structure, we achieve more accurate motion-node matching while maintaining temporal consistency across long music sequences. The proposed approach innovatively incorporates the principles of Retrieval-Augmented Generation (RAG), where the most semantically relevant motion segments are first retrieved from the motion graph, followed by diffusion-based refinement that enhances both motion quality and musical alignment.

The technical innovations of our framework include: 1) A contrastive learning architecture that learns discriminative representations for music-dance correspondence; 2) An enhanced motion graph system handles arbitrary long-term length music inputs through intelligent motion segment stitching; 3) The integration of DiT (Diffusion Transformer) ~\cite{peebles2023scalable_dit} architecture to improve the quality of generated motion sequences. Extensive experiments demonstrate that our method not only preserves the naturalness of template-based motions but also enables the creation of novel dance patterns through diffusion-based enhancement. This dual capability of leveraging existing motion knowledge while enabling creative generation represents a significant advancement in musically driven dance motion synthesis. Our approach achieves state-of-the-art performance across multiple evaluation metrics on AIST++~\cite{li2021aiFACT_aist++} and FineDance~\cite{li2023finedance} datasets, including both quantitative measures and qualitative assessments.

\section{Related Works}

Recent advances in music-conditioned dance generation have explored diverse paradigms, including contrastive learning, motion graphs, and diffusion models. Each addresses unique challenges in aligning audio and motion data. Below, we categorize existing approaches and highlight their distinctions from our proposed method.

\textbf{Contrastive Learning}. Contrastive learning has been widely adopted to bridge heterogeneous modalities. In the domain of action generation, MotionClip~\cite{tevet2022motionclip} and CLIP~\cite{wu2022wav2clip} leverage vision-language pretraining to align text and motion/image representations, while TANGO~\cite{liu2024tango} introduces a hierarchical audio-motion joint embedding space for speech-driven gesture synthesis. For audio-visual alignment, Wav2Clip~\cite{wu2022wav2clip} and Wav2Vec2~\cite{baevski2020wav2vec} demonstrate effective music-image and speech-embedding correspondence, respectively. Notably, MoMask~\cite{Guo_2024_CVPR_MoMask} employs residual cascading with discrete motion encoding to model long-term dependencies. However, these works primarily focus on text-motion (MotionClip~\cite{tevet2022motionclip}), speech-motion (TANGO~\cite{liu2024tango}), or music-image (Wav2Clip~\cite{wu2022wav2clip}) alignment. In contrast, our method explicitly addresses music-to-3D motion alignment by integrating MoMask's motion discretization with Wav2Clip's audio encoding strategy in a shared latent space. This enables unsupervised semantic correspondence without requiring paired data, a critical departure from prior methods.

\textbf{Motion Graph}. Motion graphs have been pivotal in ensuring temporal continuity in generated sequences. ChoreoMaster~\cite{chen2021choreomaster} constructs motion graphs using positional and velocity features, augmented with learned style embeddings and rhythm signatures to prevent style discontinuities. GVR~\cite{zhou2022audio_GVR} extends this to speech-driven gesture generation by incorporating SMPL~\cite{SMPL:2015} mesh IoU for node adjacency. TANGO~\cite{liu2024tango} further refines cross-modal alignment through latent feature distance metrics and employs max-connected subgraph pruning to enable infinite-length generation. HMInterp~\cite{liu2025video_HMInterp} adapts TANGO's~\cite{liu2024tango} framework to tag/description-to-dance tasks, prioritizing graph traversal cost minimization over contrastive matching. Our approach builds on TANGO's graph construction and pruning strategies but integrates contrastive learning-based node selection to ensure music-motion coherence. This hybridization of retrieval and generation principles allows seamless concatenation of motion segments while preserving rhythmic and semantic alignment.

\textbf{Diffusion Models}. Diffusion models~\cite{sohl2015deep}~\cite{ho2020denoising_diffusion} have emerged as powerful tools for motion synthesis. Early works like MotionDiffuse~\cite{zhang2024motiondiffuse} and ReMoDiffuse~\cite{zhang2023remodiffuse} establish text-to-motion generation pipelines, while MoRAG~\cite{MoRAG} partitions body segments (upper/lower body, torso) for retrieval-enhanced refinement. EDGE~\cite{tseng2023edge} and LODGE~\cite{li2024lodge} introduce controllability via music editing and coarse-to-fine generation, with LODGE++~\cite{li2024lodge++} optimizing for flexible primitives using VQ-VAE and GPT-based choreography networks. DiffDance~\cite{qi2023diffdance} and Beat-It~\cite{huang2024beat-it} further refine alignment by conditioning on contrastive audio embeddings or explicit beat loss. Our method diverges by combining multi-condition diffusion with a preprocessing network that fuses raw audio, contrastive embeddings, top-k retrieved motions, and beat annotations. This architecture enables high-fidelity generation while maintaining global music-motion synchronization, surpassing the localized refinements of ReMoDiffuse~\cite{zhang2023remodiffuse} and MoRAG~\cite{MoRAG}.

Other works explore reinforcement learning (Bailando~\cite{siyao2022bailando}, Bailando++~\cite{siyao2023bailando++}) for rhythm alignment via VQ-VAE and hybrid training strategies. While these methods emphasize temporal precision, our framework prioritizes semantic consistency through contrastive learning and hierarchical motion graph design.

By synthesizing insights from these paradigms, our work establishes a novel hybrid framework that unifies retrieval-based template fidelity with diffusion-based creative enhancement, addresses the limitations of prior methods in motion quality, diversity, and music-motion alignment.

\section{Methodology}

\begin{figure}[t]
    \centering
    \includegraphics[width=\textwidth]{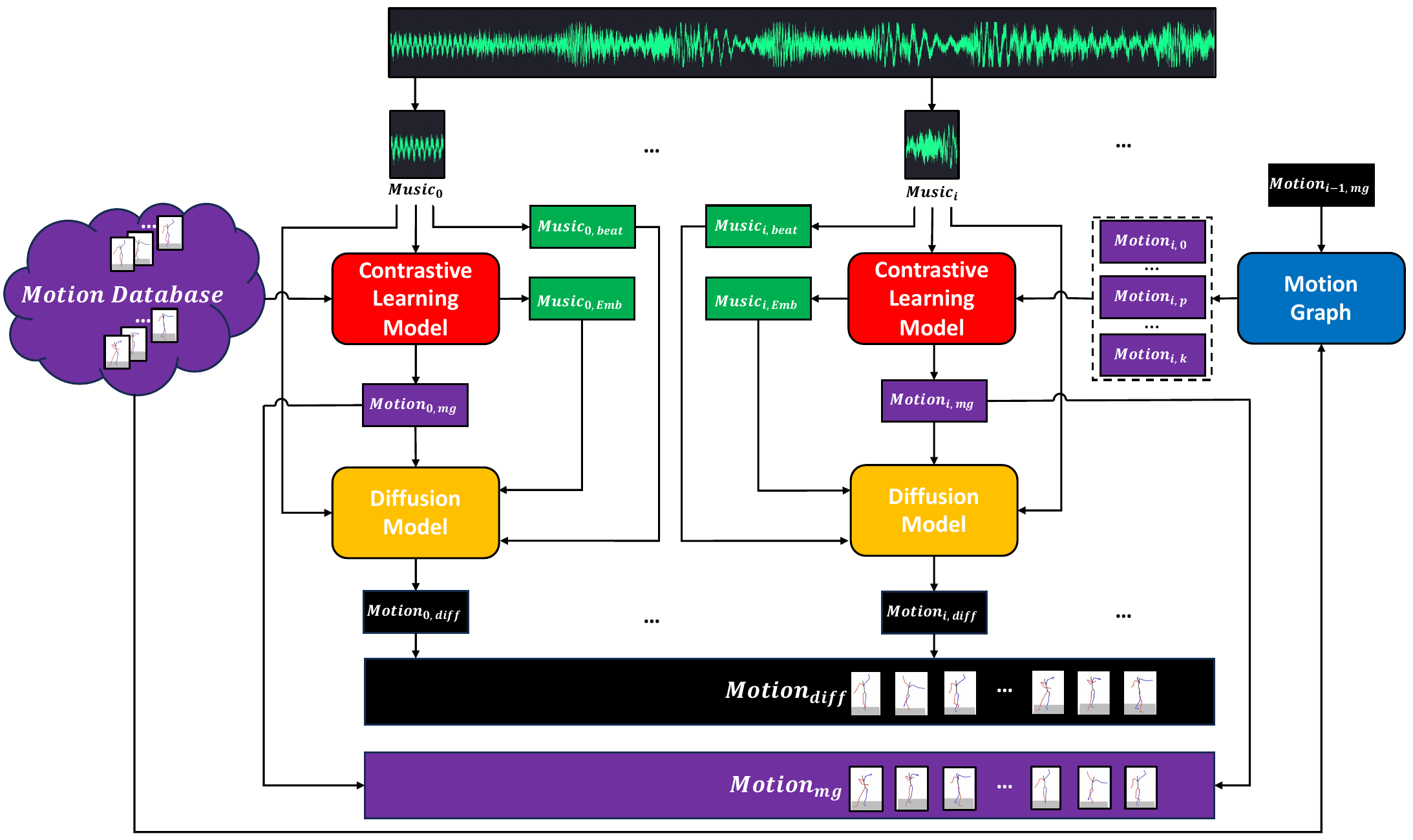} 
    \caption{The overall framework of our work. It contains three core components: the contrastive learning model, the motion graph, and the diffusion model. This integrated architecture enables the processing of arbitrarily long-term music inputs for coherent and high-quality dance motion generation.}
    \label{fig:figure_pipeline} 
\end{figure}

As illustrated in Figure~\ref{fig:figure_pipeline}, our framework consists of three main components: the Contrastive Learning Model, the Motion Graph, and the Diffusion Model. The retrieval phase, following the principles of Retrieval-Augmented Generation (RAG), is conducted between the Contrastive Learning Model and the Motion Graph to select semantically relevant motion segments. The augmentation and generation phase of RAG is then applied within the Diffusion Model to refine and enhance the retrieved motions. Consequently, our approach can be characterized as a RAG-based framework for music-to-dance generation.

\subsection{Contrastive Learning Model}

To establish the correspondence between music and motion, we employ a contrastive learning model to learn the underlying correlations from our motion database.

\begin{figure}[t]
    \centering
    \begin{minipage}[b]{0.48\textwidth}
        \centering
        \includegraphics[width=6cm]{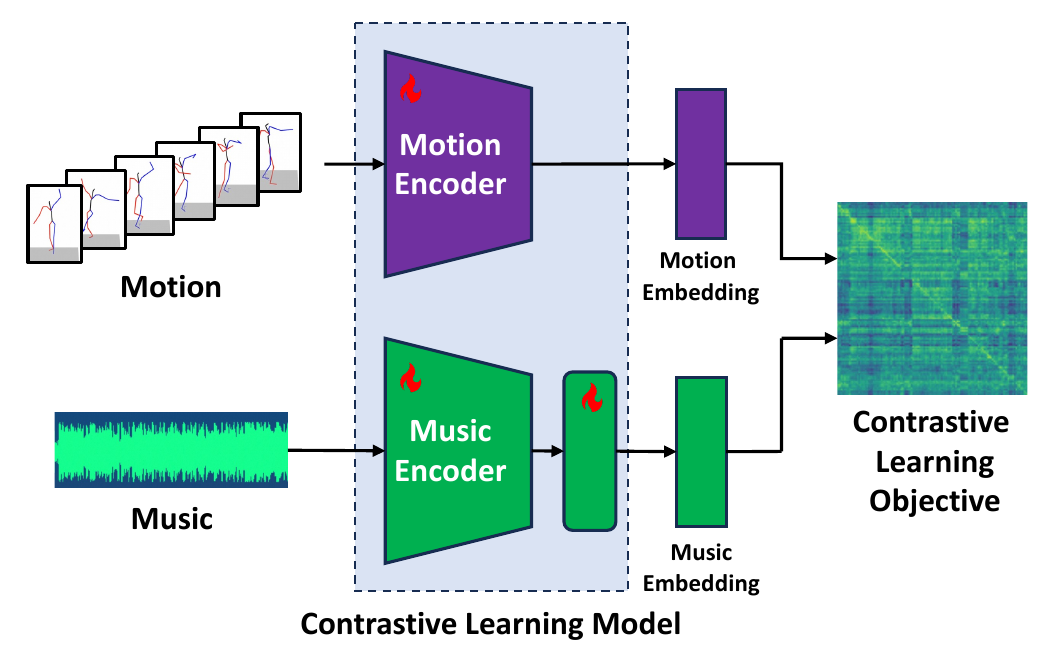}
        \caption{The contrastive learning pipeline. It contains a motion encoder, a music encoder, and an adaptive layer that follows the music encoder. All parameters in the pipeline are re-trained through the training process.}
        \label{fig:figure_contrastive_learning_model} 
    \end{minipage}
    \hspace{.15in}
    \begin{minipage}[b]{0.48\textwidth}
        \centering
        \includegraphics[width=6cm]{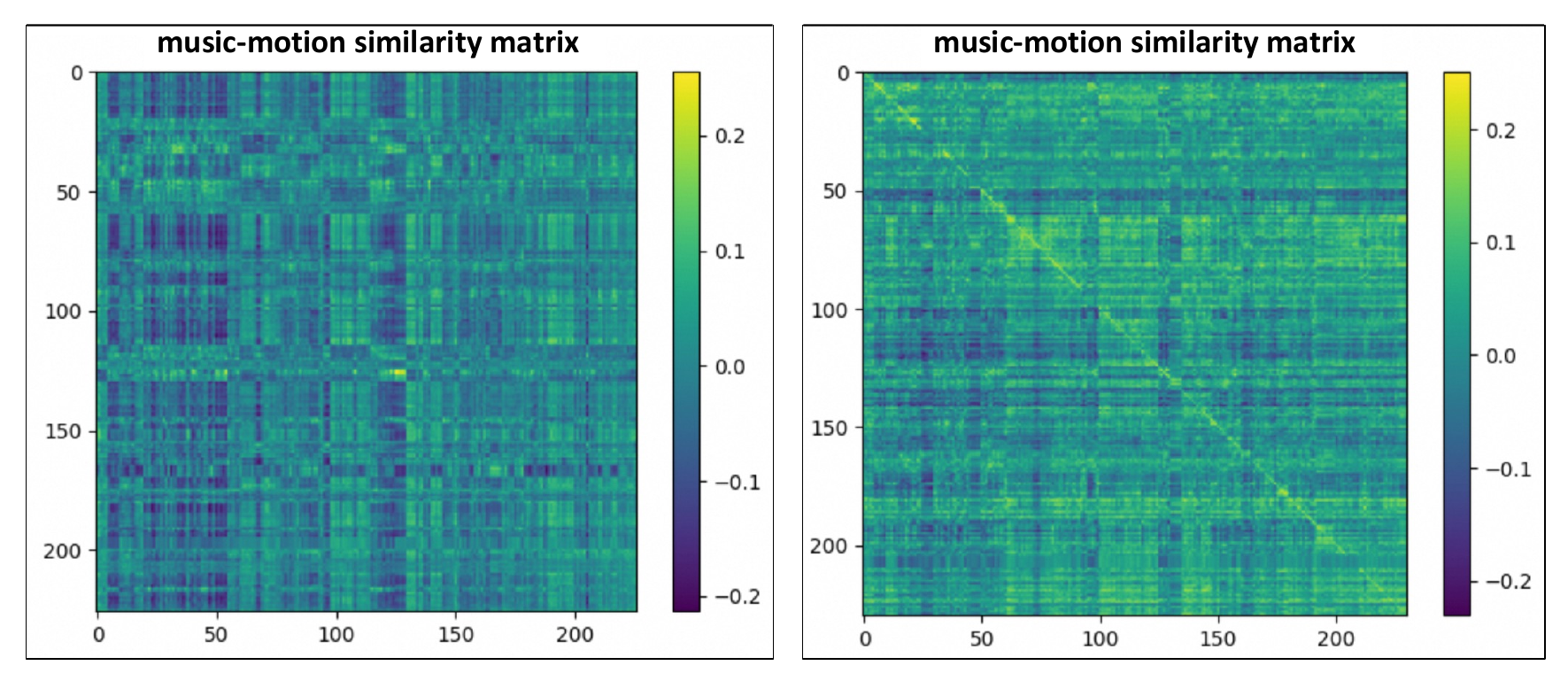}
        \caption{The music-motion similarity matrix after the contrastive learning model. The left image depicts the similarity computed directly from the raw features of motion and music, whereas the right image shows the result obtained from their embeddings after being processed by the contrastive learning model.}
        \label{fig:figure_contrastive_learning} 
    \end{minipage}
\end{figure}

As illustrated in Figure~\ref{fig:figure_contrastive_learning_model}, the contrastive learning framework consists of a motion encoder and a music encoder. The overall architecture follows a standard design similar to that in \cite{qi2023diffdance}. In our implementation, both the motion and music encoders are retrained to improve the alignment between audio and motion representations in the shared latent space.

\textbf{Motion Encoding}. We leverage the motion encoding capability provided by MoMask~\cite{Guo_2024_CVPR_MoMask} and follow the architectural settings proposed in the original work. First, we pre-train the model on the AIST++~\cite{li2021aiFACT_aist++} and FineDance~\cite{li2023finedance} datasets individually. Subsequently, we fine-tune the entire network within the contrastive learning framework to further enhance the alignment between music and motion representations.

\textbf{Music Encoding}. We follow the architectural settings proposed in \cite{wu2022wav2clip} and incorporate an adaptive layer inspired by \cite{qi2023diffdance}. While \cite{wu2022wav2clip} focuses on learning the correlation between audio and images, our task aims to model the relationship between audio and 3D motion. In our framework, we fine-tune the pre-trained music encoder and train all parameters of the adaptive layer from scratch to better align the audio and motion representations in the shared latent space.

\textbf{Music-Motion Contrastive Learning}. We learn the correlation between motion and music features using the InfoNCE loss \cite{alayrac2020self_InfoNCE}. The contrastive learning objective for the 
<music, motion> pairs is formulated as follows:
\begin{equation}
    \mathcal{L}_{i}^{m \rightarrow{d}} = 
        -log \frac{exp[{s(m_i, d_i)} / 
        \tau]}{\sum_{j=1}^{N}exp[s(m_i, d_i)/\tau]},
\end{equation}

where $m_i$ stands for the $i$-th music clip, $d_i$ stands for the $i$-th dance sequence, $\tau$ stands for a learnable temperature parameter. Figure~\ref{fig:figure_contrastive_learning_model} illustrates the overall pipeline of the training process of the contrastive learning model.

As shown in Figure~\ref{fig:figure_contrastive_learning}, the correlation between motion and music becomes significantly stronger after applying contrastive learning. The left image depicts the similarity computed directly from the raw features of motion and music, whereas the right image shows the result obtained from their embeddings after being processed by the contrastive learning model. The prominent diagonal pattern in the right image indicates that the learned representations bring the motion and music features much closer in the latent space, demonstrating the effectiveness of the contrastive learning process.

\subsection{Motion Graph}

Following a similar approach to TANGO~\cite{liu2024tango}, we construct a motion graph to establish connections among different motion segments from the motion database. The construction process comprises two main stages: graph building and graph pruning. In the first stage, all motion clips are integrated into a unified graph structure based on their compatibility in terms of position and velocity. The second stage involves pruning the graph to identify the largest connected subgraph, which ensures temporal coherence and enables the generation of arbitrarily long motion sequences.

\begin{figure}[t]
    \centering
    \includegraphics[width=\textwidth]{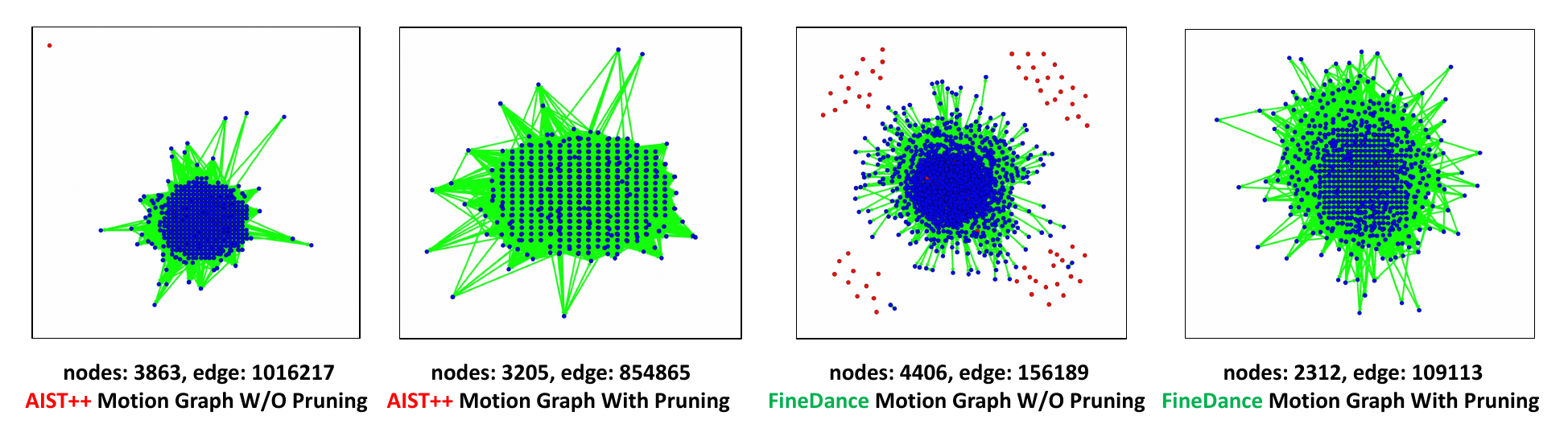} 
    \caption{The comparison of motion graph pruning results. The first and second graphs are constructed based on the AIST++~\cite{li2021aiFACT_aist++} dataset, while the third and fourth are built using the FineDance~\cite{li2023finedance} dataset. \textcolor{red}{red} points represent isolated nodes that are removed during the pruning process, whereas \textcolor{blue}{blue} points indicate connected nodes. The \textcolor{green}{green} arrows illustrate the directional relationships between adjacent, connected nodes in the pruned graph.}
    \label{fig:figure_motion_graph_prune} 
\end{figure}

\textbf{Graph Building}. The motion graph consists of nodes and edges. Each node represents a 3D motion clip, containing both positional and velocity information. Edges are constructed based on the compatibility between the position and velocity of adjacent nodes. The detailed edge-building procedure is outlined in Algorithm~\ref{alg:motion_graph_edge_building}.

\begin{center}
\begin{algorithm}[h]
    \caption{Motion Graph Edge Building Process}
    \label{alg:motion_graph_edge_building}
    \begin{algorithmic}[1]
    \Require  \\
        $ N $: number of frames for mean calculation \\
        $ T $: body joint count threshold \\
        $ \text{current node} $: sequence of joint positions/velocities \\
        $ \text{next node} $: sequence of joint positions/velocities
    \Ensure 
        Edge between nodes if motion continuity is satisfied
    
    \State $ M_p \gets \text{mean of last } N \text{ frames' positions} $
    \State $ M_v \gets \text{mean of last } N \text{ frames' velocities} $
    \State $ T_p \gets \text{current node's position differences from } M_p $
    \State $ T_v \gets \text{current node's velocity differences from } M_v $
    \State $ S_p \gets \text{position difference between last frame of current node and first frame of next node} $
    \State $ S_v \gets \text{velocity difference between last frame of current node and first frame of next node} $
    
    \If{ $ \text{each node with } Count_{S_p < T_p} \geq T $ \textbf{ and } $ Count_{S_v < T_v} \geq T $ }
        \State Add an edge between the current node and the next node
    \EndIf
    \end{algorithmic}
\end{algorithm}
\end{center}

\textbf{Graph Pruning}. Like TANGO \cite{liu2024tango}, we eliminate dead-end nodes by merging strongly connected components (SCCs). After this graph pruning process, the resulting motion graph becomes fully connected, enabling the generation of arbitrary long motion sequences.

As illustrated in Figure~\ref{fig:figure_motion_graph_prune}, the first and third graphs depict the original motion graphs constructed on the AIST++~\cite{li2021aiFACT_aist++} and FineDance~\cite{li2023finedance} datasets, respectively. The second and fourth graphs display their corresponding pruned versions. From the results, we observe that the number of nodes is reduced by 17.0\% and 47.5\%, respectively, following the pruning process. This reduction ensures that all remaining nodes form a single connected component, thereby enabling seamless traversal from any node to any other node within the graph.

\textbf{Motion Generation}. Each node in the motion graph has a high out-degree, indicating that it can transition to multiple different nodes. Directly concatenating the current node with the next one often results in a noticeable discrepancy between the last frame of the current motion segment and the first frame of the subsequent one, leading to visually jarring transitions. To address this issue, we apply smoothing techniques to ensure a more natural and continuous motion flow. Following a similar approach to \cite{zhou2022audio_GVR}, we smoothed the joint angles across adjacent connected nodes to reduce discontinuities and enhance temporal coherence.

After this process, if the input is a long-term music clip, we can generate a motion sequence of the same duration, which we refer to as $ motion_{mg} $.

\subsection{Diffusion Model}

After the motion graph process, the $ motion_{mg} $ can be used directly, but it is limited to the number of clips of the motion database, and the total motion performance is limited. Therefore, we need to augment the performance of the total motion. As the strong generation ability of the diffusion model~\cite{ho2020denoising_diffusion}, we use it to augment our music-to-dance motion.

\begin{figure}[t]
    \centering
    \includegraphics[width=\linewidth]{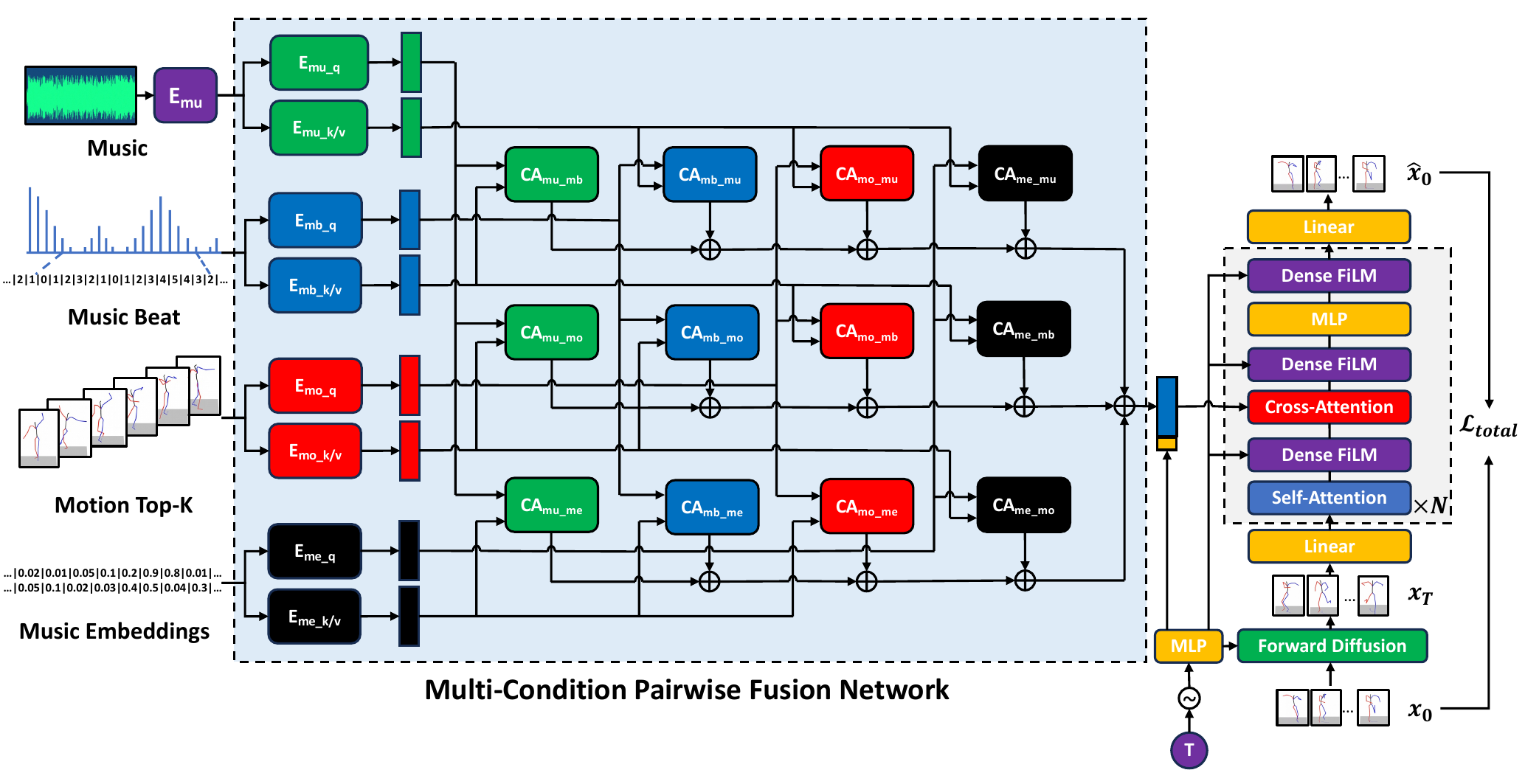} 
    \caption{The diffusion model. We propose the multi-condition pairwise fusion network to fuse the input conditions. The fused condition is then fed into the diffusion model through a cross-attention layer, enabling effective guidance of the generation process.}
    \label{fig:figure_diffusion} 
\end{figure}

An overview of our diffusion process is presented in Figure~\ref{fig:figure_diffusion}. It consists of a multi-condition fusion stage, implemented through the multi-condition pairwise fusion network, followed by the diffusion generation process, which builds upon the framework proposed in EDGE~\cite{tseng2023edge}.

\textbf{Diffusion Formulation}. Diffusion models~\cite{ho2020denoising_diffusion} define a consistent Markovian forward process that incrementally adds noise to clean sample data $ x_0^{1:L} \in q(x_0) $, along with a corresponding reverse process that gradually removes noise from corrupted samples. For brevity, we denote the entire sequence at time step $ x_t $. In the forward process, a predefined noise variance schedule $ \beta_t $ is used to control the amount of noise added at each step. The forward process can be formulated as follows:

\begin{equation}
    q(x_{1:T}|x_0) =
        \Pi_{t=1}^T q(x_t | x_{t-1}),
    q(x_t | x_{t-1}) = 
        \mathcal{N} (x_t; \sqrt{1 - \beta_t} x_{t-1}, \beta_tI).
    \label{eq:diffuse}
\end{equation}

After $ T $ steps, the input data will be transformed to the noise distribution $ q(x_T) $, which is usually a standard Gaussian distribution $ \mathcal{N}(0, I) $. In the reverse process, the noise will be removed from the noisy sample $ x_T $, and finally, the clean sample $ x_0 $ will be obtained. In our method, we need to inject other conditions to modify the generation process. Thus our object is to model the distribution $ p(x_0 | C) $ with a set of conditions $C$. Following~\cite{ho2020denoising_diffusion}, we
directly predict the clean sample $ x_0 $ from the noise distribution $ q(x_T) $ with the following objective:

\begin{equation}
    \mathcal{L}_{simple} = 
        \mathbb{E}_{x_{0} \sim q(x|C), t \sim [1, T]} 
        [\lVert x_{0}-G(x_t, t, C) \rVert _2^2].
    \label{eq:loss_simple}
\end{equation}

\textbf{Condition Model}. Similar to Beat-It~\cite{huang2024beat-it}, we also directly inject the conditions' feature into the diffusion model by the Cross-Attention module. Different from Beat-It~\cite{huang2024beat-it}, it proposes a hierarchical multi-condition fusion network to fuse the input conditions, while our approach propose a multi-condition pairwise fusion network to fuse the input conditions. 

Our method incorporates four key conditioning signals: the input music, the extracted music beat, the top-k motion candidates, and the learned music embeddings. The input music is encoded using Jukebox~\cite{dhariwal2020jukebox}. The music beat is extracted following the approach proposed in Beat-It~\cite{huang2024beat-it}~\cite{mcfee2015librosa}. The top-k motion candidates are retrieved through our previously described contrastive learning model and motion graph pipeline. Finally, the music embeddings are obtained via the contrastive learning model, which captures the semantic relationship between the audio and motion data.

The multi-condition pairwise fusion network first extracts the query ($q$), key ($k$), and value ($v$) features from each input condition. It then performs pairwise interactions by combining the query of one condition with the keys and values of all other conditions. Through this mechanism, the network enables rich and diverse fusion among multiple conditional inputs, allowing for more comprehensive and context-aware information integration during the generation process.

\textbf{Diffusion Model}. The diffusion model is built upon the architecture of ~\cite{tseng2023edge}~\cite{perez2018film}, with our primary modification focusing on the fusion of input conditional features.

\textbf{Losses}. Following ~\cite{tseng2023edge}~\cite{tevet2022human}, we also incorporate the loss terms $ \mathcal{L}_{pos} $, $ \mathcal{L}_{vel} $ and $ \mathcal{L}_{contact} $ to enforce constraints on motion position, velocity, and foot-ground contact, respectively.




\begin{align}
    \mathcal{L}_{pos} &= 
        \frac{1}{N} \sum_{i=1}^N 
        \lVert FK(x^{(i)}) - FK(\hat{x}^{(i)}) \rVert _2^2, \label{equ:loss_joint} \\ 
    \mathcal{L}_{vel} &= 
        \frac{1}{N-1} \sum_{i=1}^{N-1} 
        \lVert(x^{(i+1)} - x^{(i)}) - (\hat{x}^{(i+1)} - \hat{x}^{(i)}) \rVert _2^2, \label{equ:loss_velocity}  \\ 
    \mathcal{L}_{contact} &= 
        \frac{1}{N-1} \sum_{i=1}^{N-1} 
        \lVert  (FK(x^{(i)}) - FK(\hat{x}^{(i)})) \cdot \hat{b}^{(i)}  \rVert _2^2,\label{equ:loss_contact}
\end{align}

The total objective is as follows, while the setting of $\lambda_{pos}$, $\lambda_{vel}$, and $\lambda_{contact}$ are the same as EDGE~\cite{tseng2023edge}:
\begin{equation}
    \mathcal{L}_{total} = 
        \mathcal{L}_{simple} + 
        \lambda_{pos} \mathcal{L}_{pos} +
        \lambda_{vel} \mathcal{L}_{vel} +
        \lambda_{contact} \mathcal{L}_{contact}.
    \label{equ:loss_total}
\end{equation}

After the diffusion model processing, we obtain the augmented generated motion, denoted as $motion_{diff}$. 


\section{Experiments}
\label{section: experiments}

For the sake of simplicity, we refer to $ stage1 $ as the process that involves contrastive learning and motion graph processing, which generates $ motion_{mg} $, and $ stage2 $ as the diffusion-based refinement process that produces $ motion_{diff} $. We conduct our experiments on the dataset AIST++~\cite{li2021aiFACT_aist++} and FineDance~\cite{li2023finedance}.

\subsection{Dataset}
\textbf{AIST++}. AIST++~\cite{li2021aiFACT_aist++} is a large-scale open-source 3D dance dataset containing 1,408 music-synchronized motion sequences. It consists of 980 training sequences and 40 test sequences. The motion data is represented as 60-FPS 3D poses in SMPL~\cite{SMPL:2015} format. All experiments are conducted on the AIST++ dataset following the experimental setup outlined in \cite{siyao2022bailando}.

\textbf{FineDance}. The FineDance dataset~\cite{li2023finedance} provides 7.7 hours of 30-FPS motion data across 22 dance genres, with an average sequence length of 152.3 seconds—far exceeding AIST++’s 13.3 seconds. Captured using high-quality optical motion capture by professional dancers, it ensures both artistic quality and kinematic accuracy. We follow Lodge’s protocol~\cite{li2024lodge}, generating sequences for 20 test tracks and evaluating 1024 frames. Its long-duration and rich choreography make it a robust benchmark for music-driven motion synthesis.

\subsection{Implementation Details}

All experiments are conducted on a NVIDIA GeForce RTX 4090 GPU.

\textbf{Contrastive Learning Model}. We first pre-train the motion encoder with a batch size of 256, a maximum of 50 epochs, and a learning rate of 2e-4. The training time for this stage is approximately 4 hours. Subsequently, we train the music encoder using a batch size of 256, a maximum of 5000 epochs, and a learning rate of 1e-4. We employ AdamW~\cite{loshchilov2017decoupled_adamw} as our optimizer with the weight decay is 1e-2. The training process takes around 20 hours to complete.

\textbf{Diffusion Model}. For the diffusion model, we set the batch size as 64, the maximum number of epochs to 2000, and use a learning rate of 2e-4. We employ
Adan~\cite{xie2024adan} as the optimizer and use the Exponential Moving Average (EMA)~\cite{klinker2011exponential_ema} technique to enhance the stability of loss convergence. The model typically converges within approximately 10 hours.

\subsection{Evaluation Metrics}

We evaluate our approach using three primary metrics: FID (Fréchet Inception Distance) for motion quality, DIV (Diversity Score) for motion diversity, and BAS (Beat Alignment Score) for music-motion synchronization accuracy. These metrics are applied to assess the performance of our method on both the AIST++~\cite{li2021aiFACT_aist++} and FineDance~\cite{li2023finedance} datasets.

\textbf{Motion Quality}. This metric primarily evaluates the quality of the generated dance motion. It includes FID$_k$ and FID$_g$, which denote the Fréchet Inception Distance (FID) computed using kinematic features and geometric features, respectively. The subscripts $k$ and $g$ indicate the type of feature used in the distance calculation.

\textbf{Motion Diversity}. This metric primarily evaluates the diversity of the generated motion. It includes DIV$_k$ and DIV$_g$, which represent the diversity scores computed based on kinematic and geometric features~\cite{gopinath2020fairmotion}, respectively. We follow the approach proposed in Bailando~\cite{siyao2022bailando} to calculate these scores by measuring the average feature distances among the generated dance motions. The subscripts $k$ and $g$ denote the type of feature used for diversity estimation.

\textbf{Beat Alignment Score(BAS)}. To evaluate the accuracy of music-motion synchronization, we adopt the BAS (Beat Alignment Score) metric following the methodology in \cite{siyao2022bailando}. On the AIST++~\cite{li2021aiFACT_aist++} dataset, our approach achieves the highest score of 0.2874 in $stage1$. On the FineDance~\cite{li2023finedance} dataset, it attains a score of 0.2631 after the $stage2$ process, representing the best performance among the compared methods. The BAS is calculated using the following equation:
\begin{equation}
    BAS = \frac{1}{|B^m|} \sum_{t^m \in B^m} exp\{- \frac{min_{t^d \in B^d} \|t^d - t^m\|^2}{2\sigma^2}\}.
    \label{eq:bas}
\end{equation}

\subsection{Comparison to Existing Methods}
We primarily compare our approach with state-of-the-art music-to-dance generation methods, including Bailando++~\cite{siyao2023bailando++}, Lodge++~\cite{li2024lodge++}, and EDGE~\cite{tseng2023edge}. We do not include Beat-It~\cite{huang2024beat-it} in the comparison due to discrepancies in evaluation settings. Specifically, the reported ground-truth (GT) value for Beat-It is significantly higher than that of other methods (0.384 v.s. 0.2374), suggesting potential differences in metric computation or data normalization. Additionally, we were unable to obtain detailed evaluation protocols from the original work, which would be necessary for a fair and consistent comparison.

\textbf{Comparing on AIST++~\cite{li2021aiFACT_aist++} dataset}. As demonstrated in Table~\ref{tab:table_aistpp}, on the AIST++~\cite{li2021aiFACT_aist++} dataset, we achieve the highest BAS score of 0.2874 on $ stage1 $. Following the diffusion refinement, our method obtains a higher FID$_k$ score, which is second only to Bailando++~\cite{siyao2023bailando++} but significantly outperforms EDGE~\cite{tseng2023edge} and Lodge~\cite{li2024lodge}, both of which operate in the same long-term music-to-dance generation task. This demonstrates the effectiveness of our hybrid approach in balancing motion quality and temporal coherence.

\begin{table}
    \centering
    \caption{Compare with SOTAs on the AIST++~\cite{li2021aiFACT_aist++} dataset. The best and runner-up values are bold and underlined, respectively. $\downarrow$ means lower is better. $\uparrow$ means upper is better.}
    \label{tab:table_aistpp}
    \renewcommand
    \arraystretch{1.2}
    \begin{tabular}{cccccc} 
         \toprule
         \cmidrule{1-6}
         \multirow{2}{*}{Method} & \multicolumn{2}{c}{Motion Quality} & \multicolumn{2}{c}{Motion Diversity} & \multirow{2}{*}{BAS$\uparrow$} \\
         \cmidrule(r){2-3} \cmidrule(r){4-5}
               & FID$_{k}\downarrow$ & FID$_{g}\downarrow$ & Div$_{k}\uparrow$ & Div$_{g}\uparrow$ & \\ 
         \hline
         Ground Truth & 17.10 & 10.60 & 8.19 & 7.45 & 0.2374 \\ 
         \hline
         Li ${et\ al.}$ \cite{li2020learning} & 86.43 & 43.46 &   6.85 & 3.32 & 0.1607 \\ 
         DanceNet \cite{zhuang2022music2dance_dancenet} & 69.18 & 25.49 &   2.86 & 2.85 & 0.1430 \\
         DanceRevolution \cite{huang2020dance_DanceRevolution} & 73.42 & 25.92 & 3.52 & 4.87 & 0.1950 \\
         FACT \cite{li2021aiFACT_aist++} & 35.35 & 22.11 & 5.94 & 6.18 & 0.2209 \\ 
         Bailando \cite{siyao2022bailando} & 28.16 & \textbf{9.62} & \underline{7.83} & \underline{6.34} & 0.2332 \\
         Bailando++ \cite{siyao2023bailando++} & \textbf{17.59} & \underline{10.10} & \textbf{8.64} & \textbf{6.50} & \underline{0.2720} \\
         EDGE \cite{tseng2023edge} & 42.16 & 22.12 & 3.96 & 4.61 & 0.2334 \\
         Lodge \cite{li2024lodge} & 37.09 & 18.79 & 5.58 & 4.85 & 0.2423 \\
         \hline
         Ours($stage1$) & 30.17 & 19.80 & 5.82 & 6.07 & \textbf{0.2874} \\
         Ours($stage2$) & \underline{26.23} & 17.66 & 5.62 & 3.79 & 0.2545 \\
         \bottomrule
    \end{tabular}
\end{table}

\textbf{Comparing on FineDance~\cite{li2023finedance} dataset}. As demonstrated in Table~\ref{tab:table_finedance}, our proposed framework demonstrates competitive performance in multiple metrics compared to the existing state-of-the-art methods in the FineDance~\cite{li2023finedance} data set. Regarding motion quality, our method achieves the lowest FID$_k$ (10.51) and FID$_g$ (20.25) scores in $ stage1 $, indicating superior fidelity to ground-truth motion distributions. The $ stage2 $ maintains strong quality (FID$_k$=32.25) while achieving the highest BAS (0.2631), reflecting its ability to balance semantic alignment with musical inputs.

For motion diversity, the $ stage1 $ achieves the highest Div$_k$ (10.67), outperforming all prior methods, including the baseline Lodge++~\cite{li2024lodge++} (Div$_k$=5.53). However, the $ stage2 $ shows a slight trade-off in diversity (Div$_k$=8.94), which is still comparable to top-performing models like EDGE~\cite{tseng2023edge} (Div$_k$=8.13). Notably, both stage processing exhibit distinct strengths: $ stage1 $ excels in maintaining high-quality and diverse motion patterns, while $ stage2 $ prioritizes music-motion semantic consistency as evidenced by its best-in-class BAS.

Compared to the previous SOTA (Lodge++~\cite{li2024lodge++}), our hybrid approach combining motion graph retrieval and diffusion-based refinement achieves significant improvements in both quality (FID$_k$ reduced by 73.1\%) and semantic alignment (BAS increased by 11.6\%), demonstrating the effectiveness of integrating retrieval-augmented generation with diffusion modeling for long-term music-driven dance synthesis.

\begin{table}
    \centering
    \caption{Compare with SOTAs on the FineDance~\cite{li2023finedance} dataset. The best and runner-up values are bold and underlined, respectively. $\downarrow$ means lower is better. $\uparrow$ means upper is better.}
    \label{tab:table_finedance}
    \renewcommand\arraystretch{1.2}
    \begin{tabular}{cccccc} 
         \toprule
         \cmidrule{1-6}
         \multirow{2}{*}{Method} & \multicolumn{2}{c}{Motion Quality} & \multicolumn{2}{c}{Motion Diversity} & \multirow{2}{*}{BAS$\uparrow$} \\
         \cmidrule(r){2-3} \cmidrule(r){4-5}
               & FID$_{k}\downarrow$ & FID$_{g}\downarrow$ & Div$_{k}\uparrow$ & Div$_{g}\uparrow$ & \\ 
         \hline
         Ground Truth & / & / & 9.73 & 7.44 & 0.2120 \\ 
         \hline
         FACT \cite{li2021aiFACT_aist++} & 113.38 & 97.05 & 3.36 & \underline{6.37} & 0.1831 \\ 
         MNET \cite{kim2022brand_MNET} & 104.71 & 90.31 & 3.12 & 6.14 & 0.1864 \\
         Bailando \cite{siyao2022bailando} & 82.81 & \underline{28.17} & 7.74 & 6.25 & 0.2029 \\
         EDGE \cite{tseng2023edge} & 94.34 & 50.38 & 8.13 & \textbf{6.45} & 0.2116 \\
         Lodge \cite{li2024lodge} & 50.00 & 35.52 & 5.67 & 4.96 & 0.2269 \\
         Lodge++ \cite{li2024lodge++} & 40.77 & 30.79 & 5.53 & 5.01 & 0.2423 \\
         \hline
         Ours($stage1$) & \textbf{10.51} & \textbf{20.25} & \textbf{10.67} & 5.24 & \underline{0.2612} \\
         Ours($stage2$) & \underline{32.25} & 57.63 & \underline{8.94} & 3.75 & \textbf{0.2631} \\
         \bottomrule
    \end{tabular}
\end{table}

\section{Conclusion and Limitation}
\label{section: conclusion and limitations}

In this paper, we present a hybrid framework that combines motion graph retrieval with diffusion-based generation for long-term music-conditioned dance motion synthesis. By integrating contrastive learning, an optimized motion graph, and a DiT-based diffusion model, our method achieves superior performance in both motion quality and music alignment, as demonstrated on the AIST++~\cite{li2021aiFACT_aist++} and FineDance~\cite{li2023finedance} datasets.

Despite these promising results, our approach has certain limitations. First, the motion diversity is still constrained by the pre-built motion graph. Second, complex or ambiguous musical inputs may challenge the current alignment mechanism. Lastly, the two-stage pipeline increases computational cost, which limits real-time deployment. Future work will focus on improving efficiency, reducing dependency on large motion libraries, and enabling interactive control over generated motions.

\clearpage
\bibliographystyle{plain}
\bibliography{refs}

\end{document}